# Towards a low cost lead assay technique for drinking water using CMOS sensors

Alexander Deisting

***Abstract*—An estimated 26 million people in low- and middle-income countries are at risk of lead exposure and there is no safe threshold for lead ingestion. Radio assay methods are not easily accessible in regions at risk, therefore a low cost and easy to use sensor is desirable. Pb occurs together with traces of radioisotopes with decay energies in the range of $10$ to several $100\,\mathrm{keV}$ and beyond. Such energies are accessible in silicon sensors. We have tested a scientific CMOS (Neo 5.5 sCMOS), optimised for optical wavelengths, as $\gamma$ detector for radiation in the range of $0$ to a few $10\,\mathrm{keV}$. We find a minimal detectable $^{241}$Am decay rate of $20\,\mathrm{Bq}$ for a $\leq 1.4\,\mathrm{h}$ measurement. Optimising our analysis software will potentially enable detecting lower rates in the same measurement time. We established that the Neo 5.5 sCMOS allows to measure a spectrum of $^{241}$Am decay lines. In addition we show that it is possible to enhance the concentration of radioisotopes in water when reducing the water's volume by boiling. The knowledge gained using the scientific CMOS sensor will be transferred to commercial silicon sensors as the chips in smart phones.**

## I. Introduction

It is estimated that globally 26 million people in low- and middle-income countries are at risk of lead exposure [1] and there is no safe threshold below which lead can be considered harmless to human health [2]. Lead assay methods are often not easily available in regions where lead contaminations are frequent. Therefore, a low cost and easy to use sensor could have a huge impact to reduce lead ingestion in at-risk regions. Lead occurs together with traces of radioisotopes, including the radioactive lead isotope $^{210}$Pb. The corresponding decay energies cover the range of $10$ to several $100\,\mathrm{keV}$, which is easily accessible in silicon sensors. Measuring lead concentrations at, and below the World Health Organisation (WHO) limit for drinking water of $0.87\,\mathrm{ppb}$ [2] can be done with methods developed for Dark Matter searches, which have reached sensitivities down to $10^{-10}\,\mathrm{ppb}$ of $^{210}$Pb in acrylic [3], or $10^{-8}\,\mathrm{ppb}$ of $^{210}$Pb in water [4]. A low cost sensor will have orders of magnitude less sensitivity, which is potentially still enough to measure Pb at the WHO limit.

This contribution is on behalf of working-groups and researches of the following institutions: Centro Atómico Bariloche, Argentina; Boulby Underground Laboratory, Whitby, UK; Instituto de Ciencias Nucleares, Universidad Nacional Autónoma de México, Mexico; Instituto de Física, Universidad Nacional Autónoma de México, Mexico; Royal Holloway, University of London, UK; University College London, UK; University of Sussex, UK

This research is founded by the Science & Technology Facilities Council GCRF Foundation Award grant reference ST/R002908/1

A. Deisting is with Royal Holloway, University of London, Egham Hill, Egham, TW20 0EX, UK (e-mail: alexander.deisting@rhul.ac.uk).

## II. Experimental Set-up

We use an Oxford Instruments/Andor Neo 5.5 sCMOS sensor with $2560 \times 2160$ pixel. The set-up is hosted in a dark-box with dimensions of $244\,\mathrm{cm} \times 122\,\mathrm{cm} \times 122\,\mathrm{cm}$ (LxWxH). For a measurement, the test object – usually a radioactive source – is positioned in front of the camera and the dark box is closed. Data is written out in the FITS format and processed by our analysis software. The measurements shown in Sections III and IV are done with an $^{241}$Am source. $^{241}$Am has two possible $\alpha$ decay channels, which include the emission of $\gamma$-ray, to an excited state of $^{237}$Np, which de-excites by emitting a $\gamma$. The $\alpha$s are stopped by the camera's window, whereas the $\gamma$s reach the CMOS. The most probable $\gamma$ energies ($\varepsilon_\gamma$) are $59.5\,\mathrm{keV}$ and $26.3\,\mathrm{keV}$, liberating 16312 and 7218 electrons in the Silicon respectively.

## III. Analysis procedure

A recorded image (*frame*) is a 2D ($x/y$) matrix filled with charge values (Fig. 1a). From a series of $k$ frames we construct a pedestal value and a threshold value for each position in the 2D matrix. The pedestal value of a pixel in a certain $x/y$ position corresponds to the mean of the $k$ charge values measured by the pixel in $k$ analysed frames. The threshold for radiation detection is calculated from the respective standard deviation of the mean. Using more than one frame allows to reject charge values in the calculation which have an anomalous high charge – possibly due to the presence of ionising radiation.

Clusters are searched in each frame after the pedestals and thresholds are calculated. All charge values larger than a pixel's pedestal value summed with its threshold are considered as clusters (Fig. 1b). Doing so, adjacent pixel with charge values over threshold are grouped together as one cluster.

## IV. Results

After the cluster finding, the total charge in each cluster is calculated. Histograming the charge of the identified clusters yields the line spectrum in Figure 1c. More than four peaks can be easily identified in the spectrum – the exact matching to $^{241}$Am and $^{237}$Np lines is still work in progress.

### A. *Smallest measurable source activity*

We perform measurements with the $^{241}$Am source whilst increasing the distance between Neo sCMOS source, to test the smallest $\gamma$-ray activity measurable. The resulting spectra are normalised and background corrected.

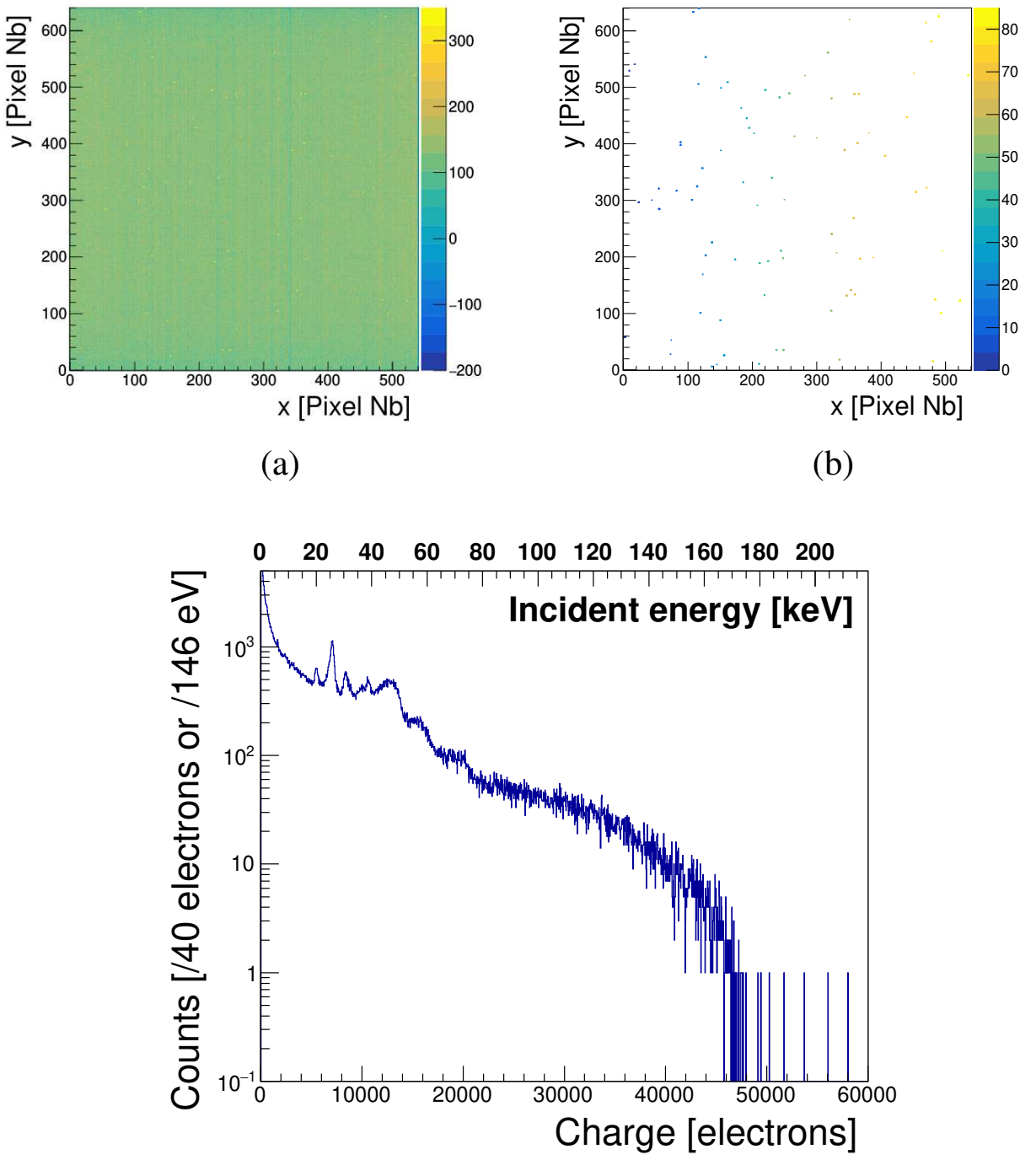

(a) (b)

(c)

Fig. 1: Measurement with an $^{241}$Am source. (a) Raw data frame (with $4 \times 4$ re-binning)as recorded by the camera. (b) Identified clusters in the raw image. (c) Spectrum of the cluster charge of several frames.

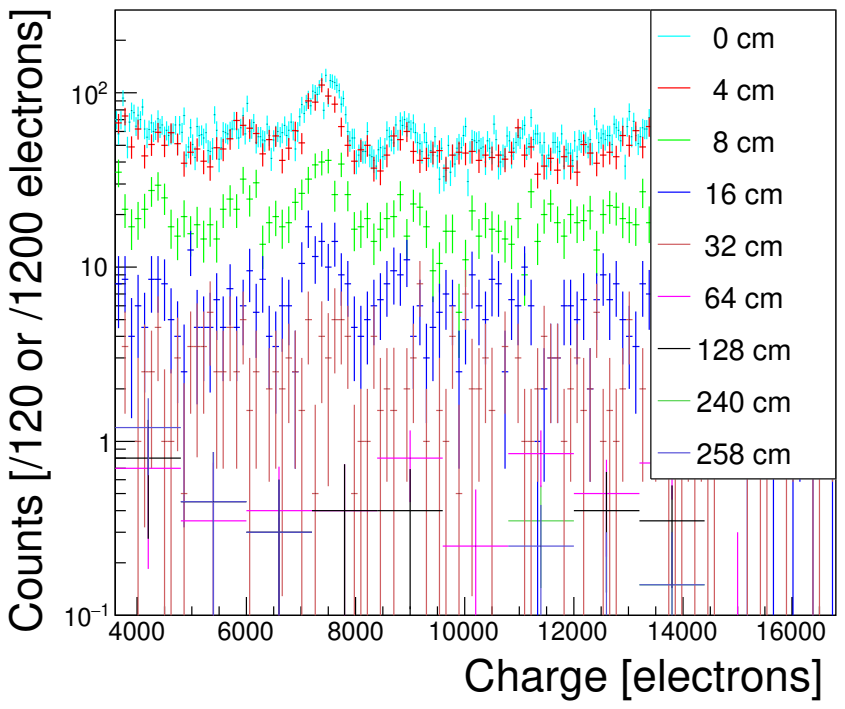

Fig. 2: Several measurements of an $^{241}$Am spectrum for increasing distance between camera and source. All data is normalised to live time.

Using the known source rate and the geometrical acceptance of the camera we find a value of just about $20\,\mathrm{Bq}$ for $5 \times 10^3\,\mathrm{s}$ measurement time – this corresponds to the $64\,\mathrm{cm}$ point in Fig. 2. The ability to perform calorimetry – *i.e.* how well peaks can be identified – deteriorates already at higher rates. Increasing the measurement time allows to measure lower rates and helps to regain calorimetry capabilities.

### B. *Enhancing sensitivity*

To improve on the minimal detectable rate either the measurement time can be increased or the source rate can be enhanced. We test concentrating natural radioisotopes present in water by boiling – keeping in mind a future application to drinking water assay. London tap water samples with different reduction factors are then assayed at the Boulby Underground Laboratory with high precision $\gamma$-ray counters. Figure 3 shows two examples, zoomed into the $^{40}$K peak. Between Fig. 3a and Fig. 3a the water has been reduced by a factor of $\sim 30$.

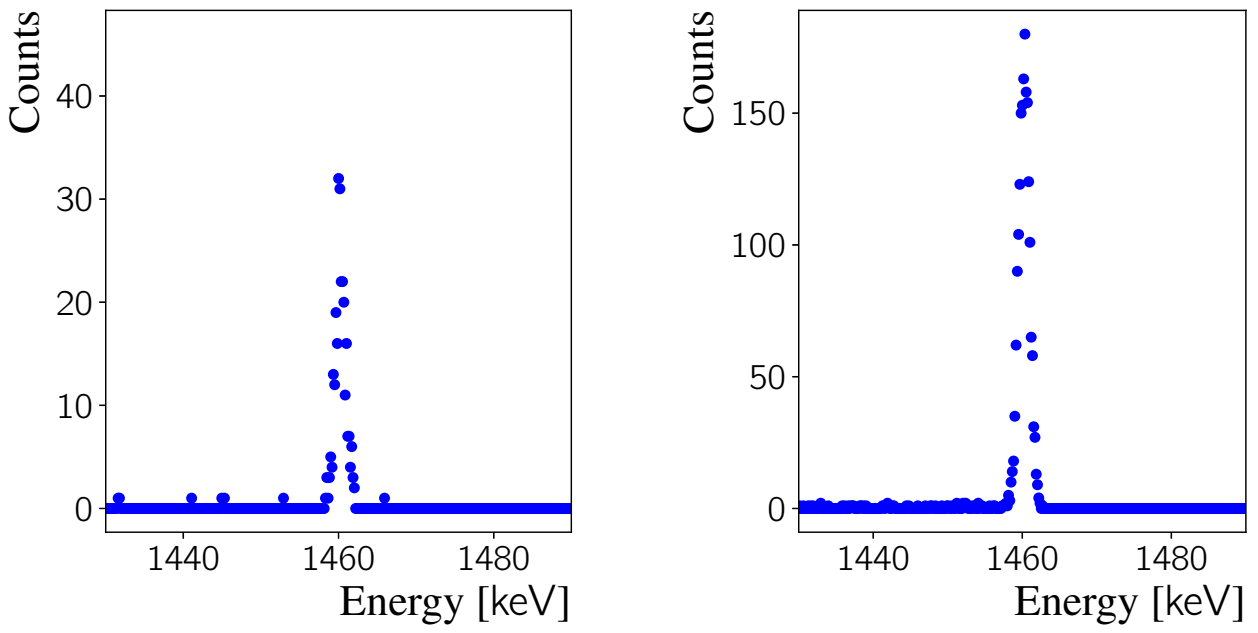

(a) (b)

Fig. 3: Two $\gamma$ spectra of reduced London tap water samples measured with $\gamma$ counters for the same measurement time, zoomed into the $^{40}$K peak. The samples have been concentrated by different reduction factors – from (a) to (b) a factor of $\sim 30$ has been applied.

## V. Conclusions & Outlook

We show that a scientific CMOS camera, built for detection of optical wavelengths, works as radiation detector for $\gamma$ radiation from about 10 to $100\,\mathrm{keV}$. With enough statistics a decay radiation peak-spectrum can be measured, which allows to identify the detected radiation. For a $\leq 1.4\,\mathrm{h}$ measurement, the limit on the detectable rate is $20\,\mathrm{Bq}$. Reducing the volume of a water sample allows to increase the source rate.
At a lead concentration of $1\,\mathrm{ppb/g}$ $H_2O$ there are $0.33 \times 10^{13}$ Pb $\mathrm{atoms/g}$. Less than that will be $^{210}$Pb, so one would expect less than $0.3\,\mathrm{decays/s}$. Measuring $0.3\,\mathrm{decays/s}$ and below is in reach when combining measurement times of a few $\mathrm{h}$ and the volume reduction technique. Improving our background correction method will bring the minimal detectable rate further down. Furthermore we are applying methods used the Neo sCMOS to a smart-phone CMOS.